\documentclass[conference]{IEEEtran}

\usepackage[utf8]{inputenc}
\usepackage[english]{babel}

\usepackage{amsthm}
\usepackage{xpatch}
\usepackage{color}
\makeatletter
\xpatchcmd{\proof}{\@addpunct{.}}{\@addpunct{:}}{}{}
\makeatother
\usepackage[abs]{overpic}
\usepackage{cite}
\usepackage{amssymb}
\usepackage{amsmath}
\usepackage{graphicx}

\usepackage[figurename=Fig.]{caption}
\usepackage{comment}
\newtheorem{theorem}{Theorem}

\newtheorem{lemma}[theorem]{Lemma}
\usepackage{enumerate}

\DeclareMathOperator*{\argmin}{arg\,min}
\renewcommand\IEEEkeywordsname{Index Terms}

\IEEEoverridecommandlockouts
\begin{document}
%
\title{Cell-Free Massive MIMO With Radio Stripes and\\ Sequential Uplink Processing}


%
\author{Zakir Hussain Shaik, Emil Bj{\"o}rnson and Erik G. Larsson \\
	Department of Electrical Engineering (ISY), Link{\"o}ping University, Link{\"o}ping, Sweden\\
	Email: \{zakir.hussain.shaik, emil.bj{\"o}rnson, erik.g.larsson\}@liu.se\thanks{This work was partially supported by the Swedish Research Council (VR) and ELLIIT.}}


\maketitle

\begin{abstract}
Cell-free Massive MIMO (mMIMO) is envisaged to be a next-generation technology beyond 5G with its high spectral efficiency and superior spatial diversity as compared to that of conventional MIMO technology. The main principle is that many distributed access points (APs) cooperate  to simultaneously serve all the users within the network without creating cell boundaries. This paper considers the uplink of a cell-free mMIMO system utilizing the radio stripe network architecture. We propose a novel sequential processing algorithm with normalized linear minimum mean square error (N-LMMSE) combining at every AP. This algorithm enables interference suppression in cell-free mMIMO while keeping the cost and front-haul requirements low. The spectral efficiency of the proposed algorithm is computed and analyzed. We conclude that it provides an attractive trade-off between low front-haul requirements and high spectral efficiency.
\end{abstract}
\vspace{-2mm}
\begin{IEEEkeywords}
	Beyond 5G, radio stripes, cell-free Massive MIMO, uplink, N-LMMSE processing, spectral efficiency.
\end{IEEEkeywords}

\vspace{-2mm}

%
\IEEEpeerreviewmaketitle

\vspace{-1.2mm}
\section{Introduction}
\vspace{-1.2mm}
Massive multiple-input multiple-output (mMIMO) is one of the big advancements in the field of wireless communications in recent years. Its high spectral efficiency (SE), beamforming gain, and reliability have made it a key physical layer technology in 5G \cite{5G_NR}.  However, it is limited by inter-cell interference, due to its cell-centric implementation. This can be overcome by so-called cell-free mMIMO which is a type of distributed mMIMO implementation with user-centric design \cite{Hien,Nayebi2016a}. In cell-free mMIMO networks, many distributed APs are connected to a central processing unit (CPU) and they jointly serve all the user equipments (UEs) within the network simultaneously. An AP can be thought of as a circuitry comprising of antenna elements and the signal processing units required to operate them, such as filters, analog-digital and digital-analog converters (ADC and DAC), etc. This kind of setup helps in performing computations locally.

The original form of cell-free mMIMO requires a dedicated front-haul and power supply to every AP \cite{Hien,Nayebi2016a}. In the uplink, each AP pre-processes the received signals and computes channel estimates, which are then sent over parallel front-haul connections to the CPU, which combines the signals. While this architecture is preferable from a communication performance perspective \cite{bjrnson2019making}, its practical adoption is questionable from a cost perspective since a huge number of long cables are needed. Hence, we need to find more practical architectures and ways to decentralize the processing.

Different techniques and algorithms for decentralizing the processing in mMIMO systems have recently been proposed \cite{8770280,8114173,7395392,7905910,8417560,sadeghi,li2019design,8891538,sanchez2019iterative,sanchez2018fully}. Prior works have considered: Fully centralized (all processing is done at the CPU) and fully distributed implementations (all processing is done at the APs, except for fusing the information at the CPU, using statistical information). The related works which focus on developing algorithms for decentralization  \cite{8770280,8114173,7395392,7905910,8417560,sadeghi,li2019design,8891538,sanchez2019iterative,sanchez2018fully} have considered daisy-chain like approach to approximate zero forcing (ZF), variants of maximum ratio (MR) processing, etc., for mMIMO \cite{8770280,8114173,7395392,7905910,8417560} and large intelligent surfaces \cite{8891538,sanchez2018fully,sanchez2019iterative}. In \cite{KarrayTree}, a tree-based architecture is proposed for mMIMO with MR and ZF but detailed signal processing techniques were not developed. 

One way to implement cell-free mMIMO is using so-called radio stripes 
 \cite{interdonato2019ubiquitous}, which are suitable for deployments in dense areas such as stadiums and malls with many APs per km$^2$. 
  In a radio stripe network, the APs are sequentially connected (i.e., using a daisy-chain architecture) and share the same cables for front-haul and power supply.\footnote{In a large cell-free mMIMO network, there will be multiple radio stripes.} Hence, there is a sequential front-haul as illustrated in Fig.~\ref{fig0}, which reduces the cabling substantially. Existing works have shown that MR combining can be computed sequentially over the front-haul in a radio stripe \cite{interdonato2019ubiquitous}, but there is no prior work that lets neighboring APs cooperate. In other words, the processing scheme does not exploit the architecture of the radio stripe.

\textbf{Contributions}:
In this paper, we propose sequential uplink processing for cell-free mMIMO based on radio stripes. The APs are pairwisely cooperating by passing around a small amount of channel state information (CSI) to enable interference suppression. Each AP computes local channel estimates and makes soft estimates of the desired signals using N-LMMSE (normalized linear mean square error) combining and then forwards the soft estimates, CSI and error statistics to the next AP, which improves the soft estimates using the available CSI. This sequential processing helps in improving the accuracy of the data estimates. This process continues sequentially until the final AP computes the final signal estimates, which are forwarded to the CPU for final decoding. The algorithm proposed in this paper differs from \cite{8891538,sanchez2018fully,sanchez2019iterative} in the following aspects: $(i)$ we take into consideration the imperfect CSI which is practical, $(ii)$ in our setup, each AP not only shares its own data estimate but also its channel estimates, error statistics to the successive AP to improve the performance in terms of SE, $(iii)$ among linear estimator we have considered is N-LMMSE which maximizes SE at each AP locally.
Besides this, although the signal processing is done sequentially in \cite{8891538,sanchez2018fully,sanchez2019iterative}, the physical topology considered is non distributive, which is different from our considered system model.    
%
%
%
%

The key aspects of this work are: $(i)$ A sequential processing framework for radio stripe networks which reduces the front-haul connections, $(ii)$ Closed-form expression for the SE, $(iii)$ SE analysis using N-LMMSE combining vectors and its comparison with centralized processing.

\textit{Notations:} Boldface lowercase letters, $\mathbf{a}$, denote column vectors and boldface uppercase letters, $\mathbf{A}$, denote matrices. The superscripts $(\cdot)^*,~(\cdot)^T,$ and $(\cdot)^H$ denote conjugate, transpose, and Hermitian transpose, respectively. The $N\times N$ identity matrix is  $\mathbf{I}_N$ and the $N\times N$ zero matrix is  $\mathbf{O}_N$. A block diagonal matrix is represented by $\mathrm{bldiag}(\mathbf{A}_1,\cdots,\mathbf{A}_N)$ with square matrices $\mathbf{A}_1,\cdots,\mathbf{A}_N$. The absolute value of a scalar and $l_2$ norm of a vector are denoted by $\vert \cdot \vert$, and $\Vert \cdot \Vert$, respectively. We denote expectation and variance by $\mathbb{E}\{\cdot\}$ and $\mathrm{Var}\{\cdot\}$, respectively. We use $\mathbf{z} \sim \mathcal{CN}\left(\mathbf{0},\mathbf{C}\right)$ to denote a multi-variate circularly symmetric complex Gaussian random vector with covariance matrix $\mathbf{C}$. We denote the probability density function (PDF) of a random variable $x$ by $f(x)$. 
\section{Radio Stripes Network Model}
\begin{figure}[t!]
	\centering
\begin{overpic}[width=0.6\columnwidth,tics=10]{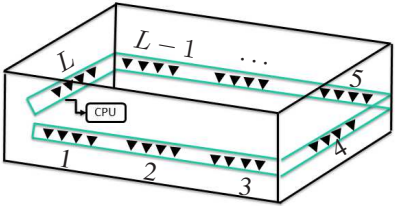}
	\put(22,15){\rotatebox{-10}{1}}
	\put(54,10){\rotatebox{-10}{2}}
	\put(90,4.2){\rotatebox{-10}{3}}
	\put(124.5,17.5){\rotatebox{30}{4}}
	\put(132,45.2){\rotatebox{-10}{5}}
	\put(90,53){\rotatebox{-10}{$\cdots$}}
	\put(50,60){\rotatebox{-10}{$L-1$}}
	\put(20,50){\rotatebox{30}{$L$}}
\end{overpic}
\vspace{-1.2mm}
\caption{Illustration of a radio stripe deployed along the walls of a room. This setup is considered in the simulation part.}
\label{fig0} \vspace{-6mm}
\end{figure}

\vspace{-1.2mm}
We consider a cell-free mMIMO radio stripe network comprising of $L$ APs, each equipped with $N$ antennas. The central processing unit (CPU) is located at the end of the stripe AP $L$, so the front-haul connections goes from AP $1$ - AP $2$ - AP $3$ - $\cdots$ - AP $L$ - CPU as shown in the Fig \ref{fig0}.
There are $K$ single antenna user equipments (UEs) distributed arbitrarily in the network and the channel between AP $l$ and UE $k$ is denoted by $\mathbf{h}_{kl} \in \mathbb{C}^{N}$. We consider the block fading channel model with coherence block length of $\tau_c$ channel uses. In each such block an independent realization is drawn from a correlated Rayleigh fading distribution as
%
\vspace{-1.2mm}
\begin{equation}
	\mathbf{h}_{kl} \sim \mathcal{CN} \left(\mathbf{0},\mathbf{R}_{kl}\right),
\end{equation}
where $\mathbf{R}_{kl} \in \mathbb{C}^{N \times N}$ is the spatial covariance matrix, which attributes the channel spatial correlation characteristics. The large-scale fading coefficient describing the shadowing and pathloss is given by $\beta_{kl}\triangleq \mathrm{tr}\left(\mathbf{R}_{kl}\right)/N$. Spatial covariance matrices $\{\mathbf{R}_{kl}\}$ are assumed to be known.

This paper studies an uplink scenario, which consists of $\tau_p$ channel uses for pilots transmission to estimate channel and $\tau_c - \tau_p$ channel uses for payload data. Both phases are described in detail below.
\vspace{-2mm}
\subsection{Channel Estimation}
We assume there are $\tau_p$ mutually orthogonal $\tau_p$-length pilot vector signals $\boldsymbol{\phi}_1,~ \boldsymbol{\phi}_2,~ \cdots,~ \boldsymbol{\phi}_{\tau_p}$ with $\Vert \boldsymbol{\phi}_{k} \Vert^2=\tau_p$, which are used for channel estimation. For the case where $K>\tau_p$, more than one UE is assigned the same pilot and hence causing so called pilot contamination. We let the pilot assigned to UE $k$, for $ k=1,\cdots, K$, to be indexed as $t_k = 1,\cdots,\tau_p$ and the set $\mathcal{S}_k=\{ i : t_i = t_k\}$ accounts for those UEs which are assigned the same pilot as that of UE $k$.
The received signal $\mathbf{Z}_l \in \mathbb{C}^{N \times \tau_p}$ at AP $l$ is
\vspace{-2mm}
\begin{equation}
	\mathbf{Z}_l = \sum_{i=1}^{K} \sqrt{p_i}\mathbf{h}_{il}\boldsymbol{\phi}_{t_i}^T + \mathbf{N}_l,
	\vspace{-2mm}
\end{equation}
where $p_i \geq 0$ is the transmit power of UE $i$, $\mathbf{N}_l \in \mathbb{C}^{N \times \tau_p}$ is the noise at the receiver modeled with independent entries distributed as $\mathcal{CN}\left(0,\sigma^2\right)$ with $\sigma^2$ being the noise power. Accordingly, the MMSE estimate \cite{mMIMObookEmil} $\widehat{\mathbf{h}}_{kl}\in \mathbb{C}^{N\times1}$ is given by
\begin{equation}
\widehat{\mathbf{h}}_{kl} = \sqrt{p_k \tau_p}\mathbf{R}_{kl}\boldsymbol{\Psi}_{t_k l}^{-1}\mathbf{z}_{t_k l},
\vspace{-5mm}
\end{equation}
where
\begin{align}
\mathbf{z}_{t_k l} &= \mathbf{Z}_l\boldsymbol{\phi}_{t_k}^*/\sqrt{\tau_p}\notag\\&= \sum_{i \in \mathcal{S}_k}\sqrt{p_i\tau_p}\mathbf{h}_{il} + \mathbf{n}_{t_k l},\\
\boldsymbol{\Psi}_{t_k l} & = \mathbb{E}\{\left(\mathbf{z}_{t_k l}-\mathbb{E}\{\mathbf{z}_{t_k l}\}\right)\left(\mathbf{z}_{t_k l}-\mathbb{E}\{\mathbf{z}_{t_k l}\}\right)^H\}\notag\\ & = \sum_{i \in \mathcal{S}_k} \tau_p p_i\mathbf{R}_{il} + \sigma^2 \mathbf{I}_N
\end{align}
is the despreaded signal and its covariance matrix, respectively. Here, $\mathbf{n}_{t_k l} \triangleq \mathbf{N}_l {\boldsymbol{\phi}_{\tau_k}^*}/{\sqrt{\tau_p}} \sim \mathcal{CN}\left(\mathbf{0},\sigma^2\mathbf{I}_{N}\right)$ is the effective noise. An important consequence of MMSE estimation is the statistical independence of the estimate $\widehat{\mathbf{h}}_{kl} \sim \mathcal{CN}(\mathbf{0},\widehat{\mathbf{R}}_{kl})$ and the estimation error $\widetilde{\mathbf{h}}_{kl} = \mathbf{h}_{kl} -\widehat{\mathbf{h}}_{kl} \sim \mathcal{CN}(\mathbf{0},\widetilde{\mathbf{R}}_{kl})$ with
\begin{equation}
\begin{aligned}[b]
\widehat{\mathbf{R}}_{kl} & = \mathbb{E}\left\{\left(\widehat{\mathbf{h}}_{kl}-\mathbb{E}\{\widehat{\mathbf{h}}_{kl}\}\right)\left(\widehat{\mathbf{h}}_{kl}-\mathbb{E}\{\widehat{\mathbf{h}}_{kl}\}\right)^H\right\}\\
&=p_k\tau_p\mathbf{R}_{kl}\boldsymbol{\Psi}_{t_k l}^{-1}\mathbf{R}_{kl},
\end{aligned}
\vspace{-4mm}
\end{equation}
\begin{equation}
\begin{aligned}[b]
\widetilde{\mathbf{R}}_{kl} & = \mathbb{E}\left\{\left(\widetilde{\mathbf{h}}_{kl}-\mathbb{E}\{\widetilde{\mathbf{h}}_{kl}\}\right)\left(\widetilde{\mathbf{h}}_{kl}-\mathbb{E}\{\widetilde{\mathbf{h}}_{kl}\}\right)^H\right\}\\ & = \mathbf{R}_{kl} -\widehat{\mathbf{R}}_{kl}.
\end{aligned}
\end{equation}
\subsection{Uplink Payload Transmission}
During the uplink payload transmission, the received signal $\mathbf{y}_l \in \mathbb{C}^N$ at AP $l$ is given by 
\begin{equation}\label{ULrecSig}
\mathbf{y}_l = \sum_{i=1}^{K}\mathbf{h}_{il}s_i + \mathbf{n}_l,
\vspace{-1mm}\end{equation}
where $s_i \sim \mathcal{CN}\left(0,p_i\right)$ is the payload signal transmitted by UE $i$ with power $p_i$ and $\mathbf{n}_l \sim \mathcal{CN}\left(\mathbf{0},\sigma^2 \mathbf{I}_N\right)$ is the independent receiver noise vector at the AP $l$.
\section{Sequential Processing}
In this section, we describe the operation of the sequential radio stipes network. All APs are pre-ordered as AP $1$ - AP $2$ - AP $3$ - $\cdots$ - AP $L$ - CPU. First the AP $1$, computes the local soft estimates $\{\widehat{s}_{k1}\}$ by using their local combining vectors $\{\mathbf{v}_{k1}\}$. Besides these, effective scalar channels and statistics of the error incurred in the effective channels are also computed. This information is shared to AP $2$ which makes use of it as the side information and computes its respective local soft estimates using its local combining vector and also effective scalar channel estimate and statistics of the error occurred. Then AP $2$ forwards this as a side information to AP $3$ and these procedure continues sequentially till AP $L$. AP $L$ forwards its computed information to CPU.

Let $\mathbf{v}_{k1} \in \mathbb{C}^N$, with $\Vert \mathbf{v}_{k1} \Vert^2=1$, be the unit norm local combining vector that AP $1$ selects for estimating the signal $s_k$ sent by UE $k$. The combining vector $\mathbf{v}_{k1}$ is designed based on the side information $\{\Omega_{i1}=\{\widehat{\mathbf{h}}_{i1},\widetilde{\mathbf{R}}_{i1}\}:i=1,\cdots,K\}$ it has. Then, its local soft estimate $\widehat{s}_{k1}$ of $s_k$ is given by
\begin{equation}\label{l1_softEst}
\begin{aligned}[b]
\widehat{s}_{k1} = \mathbf{v}_{k1}^H \mathbf{y}_1 & = \sum_{i=1}^{K} \mathbf{v}_{k1}^H \mathbf{h}_{i1}s_i + \mathbf{v}_{k1}^H \mathbf{n}_1= \sum_{i=1}^{K}g_{{i_k}1} s_i + n_{k1},
\end{aligned}
\vspace{-2mm}
\end{equation}
where
\vspace{-2mm}
\begin{align}
g_{{i_k}1}  \triangleq \mathbf{v}_{k1}^H \mathbf{h}_{i1},\ \ 
n_{k1}\triangleq \mathbf{v}_{k1}^H\mathbf{n}_1,
\end{align}
is the effective scalar channels and effective noise, respectively.  As AP 1 doesn't have the complete knowledge of the channels $\{\mathbf{h}_{i1}:i=1,\cdots,K\}$, it cannot send the effective scalar channels $\{g_{{i_k}1}:i,k=1,\cdots,K\}$, but can send only its estimates $\{\widehat{g}_{{i_k}1}\triangleq \mathbf{v}_{k1}^H \widehat{\mathbf{h}}_{i1}:i,k =1,\cdots,K\}$ to AP 2. During this process the errors incurred in the effective channels are given by $\{\widetilde{g}_{{i_k}1}\triangleq \mathbf{v}_{k1}^H \widetilde{\mathbf{h}}_{i1}: i,k =  1,2,\cdots,K\}$. The distributions of the effective noise $n_{k1}$, and the error in the effective scalar channel $\widetilde{g}_{{i_k}1}$ conditioned on the channel estimates  $\left\{\Omega_{j1}:j=1,\cdots,K\right\}$ respectively are given by
\vspace{-1.2mm}
\begin{align}
f\left(n_{k1}|\{\Omega_{j1}\}\right)&= \mathcal{CN}\left(0,\sigma^2\right),\\
f\left(\widetilde{g}_{{i_k}1}|\{\Omega_{j1}\}\right)&=\mathcal{CN}\left(0,\widetilde{\psi}_{{i_k}1}\right),
\end{align}
\vspace{-2mm}
where
\begin{align}
\widetilde{\psi}_{{i_k}1} &\triangleq \mathbf{v}_{k1}^H  \widetilde{\mathbf{R}}_{i1} \mathbf{v}_{k1}.
\end{align}
It can be observed that the distribution of error in the effective scalar channel $\widetilde{g}_{{i_k}1}$ is only dependent on the the quantity $\widetilde{\psi}_{{i_k}1}$ and not on the entire information $\{\Omega_{j1}\}$ i.e.,
\begin{equation}\label{gtilde_1_psi1}
f\left(\widetilde{g}_{{i_k}1}|\widetilde{\psi}_{{i_k}1}\right)=f\left(\widetilde{g}_{{i_k}1}|\{\Omega_{j1}\}\right)=\mathcal{CN}\left(0,\widetilde{\psi}_{{i_k}1}\right).
\end{equation}
These quantities will be useful in the later analysis. Finally, the AP 1 transmits to AP 2 the following information: 
\begin{enumerate}[(i)]
	\item soft estimates $\{\widehat{s}_k\}$ of $s_k$,
	\item effective scalar channels estimates $\left\{\widehat{g}_{{i_k}1}\right\}$, and
	\item effective channel errors variances $\{\widetilde{\psi}_{{i_k}1}\}, ~\forall i,k \in \{1,2,\cdots,K\}$.
\end{enumerate}

For the same uplink transmission, AP 2 creates an augmented received signal for estimating $s_{k}$ using \eqref{ULrecSig} and \eqref{l1_softEst} as
\begin{equation}
\begin{bmatrix}
\mathbf{y}_2\\
\widehat{s}_{k1}
\end{bmatrix}
=
\sum_{i=1}^{K}\begin{bmatrix}
\mathbf{h}_{i2}\\
g_{{i_k}1}
\end{bmatrix} s_i
+
\begin{bmatrix}
\mathbf{n}_{2}\\
n_{k1}
\end{bmatrix}.
\end{equation}
Then AP 2 creates a soft estimate $\widehat{s}_{k2}$ of ${s}_{k}$ using the combining vector $\mathbf{v}_{k2} \in \mathbb{C}^{(N+1)}$, with $\Vert \mathbf{v}_{k2} \Vert^2=1$, which is designed based on its side information
\begin{equation}\label{sideInfo_l2}
\left\{\Omega_{i2}=\left\{\widehat{\mathbf{h}}_{i2},\widetilde{\mathbf{R}}_{i2},\widehat{g}_{{i_k}1},\widetilde{\psi}_{{i_k}1}\right\}:i=1,\cdots,K\right\}.
\end{equation}
The soft estimate $\widehat{s}_{k2}$ is given as
\begin{equation}
\begin{aligned}[b]
\widehat{s}_{k2} = \mathbf{v}_{k2}^H 
\begin{bmatrix} \mathbf{y}_2\\
\widehat{s}_{k1}
\end{bmatrix} 
& =
\sum_{i=1}^{K}\mathbf{v}_{k2}^H\begin{bmatrix}
\mathbf{h}_{i2}\\
g_{{i_k}1}
\end{bmatrix} s_i
+
\mathbf{v}_{k2}^H\begin{bmatrix}
\mathbf{n}_{2}\\
n_{k1}
\end{bmatrix}\\
& = \sum_{i=1}^{K}g_{{i_k}2} s_i + n_{k2},
\end{aligned}
\end{equation}
where
\begin{align}
g_{{i_k}2}  \triangleq \mathbf{v}_{k2}^H\begin{bmatrix}
\mathbf{h}_{i2}\\
g_{{i_k}1}
\end{bmatrix},\ \ 
n_{k2}\triangleq\mathbf{v}_{k2}^H\begin{bmatrix}
\mathbf{n}_{2}\\
n_{k1}
\end{bmatrix}\label{pdf_n_l2},
\end{align}
denote the effective scalar channels and effective noise respectively at AP 2. It can be observed that normalization of combining vectors ensures that noise variance is constant throughout the sequential process. This avoids noise amplification and computation of new noise variance at each stage of the process. Next AP 2 sends its soft estimates $\{\widehat{s}_{k2}:k=1,\cdots,K\}$ and the estimates of the effective scalar channels
\begin{equation}
\left\{\widehat{g}_{{i_k}2}\triangleq \mathbf{v}_{k2}^H \begin{bmatrix}
\widehat{\mathbf{h}}_{i2}\\
\widehat{g}_{{i_k}1}
\end{bmatrix}:~i,k = 1,2,\cdots,K\right\}
\end{equation}
to AP 3. Thereby incurring the errors given by 
\begin{equation}\label{pdf_gtilde_l2}
\left\{\widetilde{g}_{{i_k}2}\triangleq \mathbf{v}_{k2}^H \begin{bmatrix}
\widetilde{\mathbf{h}}_{i2}\\
\widetilde{g}_{{i_k}1}
\end{bmatrix}:~i,k = 1,2,\cdots,K\right\}.
\end{equation}
It can be noted from \eqref{gtilde_1_psi1} and \eqref{sideInfo_l2} that
\begin{align}
f\left(\widetilde{g}_{{i_k}1}|\{\Omega_{j2}\}\right)&=\mathcal{CN}\left(0,\widetilde{\psi}_{{i_k}1}\right).\label{pdf_gtilde_l1_psitilde}
\end{align}
From \eqref{sideInfo_l2}, \eqref{pdf_n_l2}, \eqref{pdf_gtilde_l2} and, \eqref{pdf_gtilde_l1_psitilde} the distributions of the effective noise $n_{k2}$, and the error in the effective scalar channel $\widetilde{g}_{{i_k}2}$ conditioned on the side information at AP 2  $\left\{\Omega_{j2}:j=1,\cdots,K\right\}$ respectively are given by
\begin{align}
f\left(n_{k2}|\{\Omega_{j2}\}\right)&= \mathcal{CN}\left(0,\sigma^2\right),\\
f\left(\widetilde{g}_{{i_k}2}|\{\Omega_{j2}\}\right)&=\mathcal{CN}\left(0,\widetilde{\psi}_{{i_k}2}\right),
\end{align}
where
\begin{align}\label{var_psitilde_l2}
\widetilde{\psi}_{{i_k}2} &\triangleq \mathbf{v}_{k2}^H  \left[\mathrm{bldiag}\left(\widetilde{\mathbf{R}}_{i2},\widetilde{\psi}_{{i_k}1}\right)\right] \mathbf{v}_{k2}.
\end{align}
Here we made use of \eqref{var_ctilde} from the appendix for the derivation of \eqref{var_psitilde_l2}. AP 2 also sends its effective channel errors variances $\{\widetilde{\psi}_{{i_k}2}: i,k = 1,2,\cdots,K\}$ to AP 3.
With similar reasoning used in \eqref{gtilde_1_psi1} for AP 1, it can be noted that the distribution of $\widetilde{g}_{{i_k}2}$ is only dependent on $\widetilde{\psi}_{{i_k}2}$ i.e.,
\begin{equation}\label{gtilde_2_psi2}
f\left(\widetilde{g}_{{i_k}2}|\widetilde{\psi}_{{i_k}2}\right)=f\left(\widetilde{g}_{{i_k}2}|\{\Omega_{j2}\}\right)=\mathcal{CN}\left(0,\widetilde{\psi}_{{i_k}2}\right).
\end{equation}

On similar lines, AP $l$ for $l \in \{2,\cdots,L\}$ creates an augmented received signal for estimating $s_{k}$ as
\begin{equation}
\begin{bmatrix}
\mathbf{y}_l\\
\widehat{s}_{k(l-1)}
\end{bmatrix}
=
\sum_{i=1}^{K}\begin{bmatrix}
\mathbf{h}_{il}\\
g_{{i_k}(l-1)}
\end{bmatrix} s_i
+
\begin{bmatrix}
\mathbf{n}_{l}\\
n_{k{(l-1)}}
\end{bmatrix},
\end{equation}
where $n_{k{(l-1)}}$ is the effective noise from the AP $(l-1)$.
Then AP $l$ designs the combining vector $\mathbf{v}_{kl} \in \mathbb{C}^{(N+1)}$, with $\Vert \mathbf{v}_{k2} \Vert^2=1$, based on its side information
\begin{equation}\label{sideInfo_l}
\left\{\Omega_{il}=\left\{\widehat{\mathbf{h}}_{il},\widetilde{\mathbf{R}}_{il},\widehat{g}_{{i_k}(l-1)},\widetilde{\psi}_{{i_k}(l-1)}\right\}:i=1,\cdots,K\right\}.
\end{equation}
Using the combining vector AP $l$ creates the soft estimate
\begin{equation}
\begin{aligned}[b]
\widehat{s}_{kl} = \mathbf{v}_{kl}^H 
\begin{bmatrix} \mathbf{y}_l\\
\widehat{s}_{k{(l-1)}}
\end{bmatrix} 
& =
\sum_{i=1}^{K}\mathbf{v}_{kl}^H\begin{bmatrix}
\mathbf{h}_{il}\\
g_{{i_k}{(l-1)}}
\end{bmatrix} s_i
+
\mathbf{v}_{kl}^H\begin{bmatrix}
\mathbf{n}_{l}\\
n_{k{(l-1)}}
\end{bmatrix}\\
& = \sum_{i=1}^{K}g_{{i_k}l} s_i + n_{kl},
\end{aligned}
\end{equation}
where
\begin{align}
g_{{i_k}l} \triangleq \mathbf{v}_{kl}^H\begin{bmatrix}
\mathbf{h}_{il}\\
g_{{i_k}(l-1)}
\end{bmatrix},\ \ 
n_{kl} \triangleq\mathbf{v}_{kl}^H\begin{bmatrix}
\mathbf{n}_{l}\\
n_{k(l-1)}
\end{bmatrix},
\end{align}
are the effective scalar channels and noise respectively at AP $l$. Then the AP $l$ sends its soft estimate $\{\widehat{s}_{kl}:k=1,\cdots,K\}$ and the estimate of effective scalar channels
\begin{equation}
\left\{\widehat{g}_{{i_k}l}\triangleq \mathbf{v}_{kl}^H \begin{bmatrix}
\widehat{\mathbf{h}}_{il}\\
\widehat{g}_{{i_k}(l-1)}
\end{bmatrix}:~i,k = 1,2,\cdots,K\right\}
\end{equation}
to AP $(l+1)$. Thereby incurring the errors given by 
\begin{equation}\label{pdf_gtilde_l}
\left\{\widetilde{g}_{{i_k}l}\triangleq \mathbf{v}_{kl}^H \begin{bmatrix}
\widetilde{\mathbf{h}}_{il}\\
\widetilde{g}_{{i_k}(l-1)}
\end{bmatrix}:~i,k = 1,2,\cdots,K\right\}.
\end{equation}
With similar arguments given for AP 2, it can be shown that
\begin{align}
f\left(\widetilde{g}_{{i_k}(l-1)}|\{\Omega_{jl}\}\right)&=\mathcal{CN}\left(0,\widetilde{\psi}_{{i_k}(l-1)}\right).\label{pdf_gtilde_l_psitilde}
\end{align}
The distributions of the effective noise $n_{kl}$, and the error in the effective scalar channel $\widetilde{g}_{{i_k}l}$ conditioned on the channel estimates  $\left\{\Omega_{jl}:j=1,\cdots,K\right\}$ respectively using \eqref{sideInfo_l}, \eqref{pdf_gtilde_l}, and \eqref{pdf_gtilde_l_psitilde}  are given by
\begin{align}
f\left(n_{kl}|\{\Omega_{jl}\}\right)&= \mathcal{CN}\left(0,\sigma^2\right)\label{n__Dist_l},\\
f\left(\widetilde{g}_{{i_k}l}|\{\Omega_{jl}\}\right)&=\mathcal{CN}\left(0,\widetilde{\psi}_{{i_k}l}\right)
\end{align}
where
\begin{align}
\widetilde{\psi}_{{i_k}l} &\triangleq \mathbf{v}_{kl}^H  \left[\mathrm{bldiag}\left(\widetilde{\mathbf{R}}_{il},\widetilde{\psi}_{{i_k}(l-1)}\right)\right] \mathbf{v}_{kl}.
\end{align}
AP $l$ also sends $\{\widetilde{\psi}_{{i_k}l}: i,k = 1,2,\cdots,K\}$ to AP $(l+1)$, for the reason explained earlier in the AP 1 and AP 2 cases.
Finally, the AP $L$ forwards $\{\widehat{s}_{iL}\}$ and the side information
\begin{equation}
\{\Omega_{{i_k}\mathrm{CPU}}=\{\widehat{g}_{{i_k}L},\widetilde{\psi}_{{i_k}L}\}\}, \forall~i,k=1,\cdots,K
\end{equation} to the CPU for further processing. The received signal at the CPU is given by
\begin{equation}
\begin{aligned}[b]
\widehat{s}_{kL} & = \sum_{i=1}^{K} \mathbf{v}_{kL}^H \mathbf{h}_{iL}s_i + \mathbf{v}_{kL}^H \mathbf{n}_L\\
& = \sum_{i=1}^{K}\widehat{g}_{{i_k}L} s_i + \sum_{i=1}^{K}\widetilde{g}_{{i_k}L} s_i + n_{kL}.
\end{aligned}
\end{equation}

\begin{lemma}
The achievable SE of UE $k$ is
\begin{equation}\label{Spect_Eff}
\mathrm{SE}_k = \left(1 - \frac{\tau_p}{\tau_c}\right)\mathbb{E}\left\{\mathrm{log}_2 \left(1 + \mathrm{SINR}_k\right)\right\},
\end{equation}
with the effective $\mathrm{SINR}_k$ given by
\begin{equation}
\mathrm{SINR}_k = \frac{p_k\vert \widehat{g}_{{k_k}L} \vert^2}{\sum_{i=1,i\neq k}^{K}p_i\vert \widehat{g}_{{i_k}L} \vert^2 + \sum_{i=1}^{K}p_i\widetilde{\psi}_{{i_k}L}+\sigma^2},
\end{equation}
where the expectation is with respect to the effective channel estimates.
\end{lemma}
The proof of Lemma $1$ is omitted due to space limitation.

\section{Combining Vectors}
The choice of combining vector plays a crucial role in the performance of the system under consideration. In this work we take normalized LMMSE (N-LMMSE) for analysis. The choice for N-LMMSE at each AP stems from the motivation that it maximizes the SE at each AP in the sequential processing. Combining vectors $\{\mathbf{v}_{kl}:\Vert \mathbf{v}_{kl} \Vert^2 = 1, k = 1,\cdots, K,~l=1,\cdots,L\}$ are given below.

For AP $1$, N-LMMSE receiver $\mathbf{v}_{k1}^{\mathrm{N-LMMSE}} \in \mathbb{C}^{N}$ is given by
\begin{equation}
\mathbf{v}_{k1}^{\mathrm{N-LMMSE}} = \frac{\argmin_{\{\mathbf{v}_{k1}\}} \quad \mathbb{E}\left\{\vert s_{k1} - \widehat{s}_{k1}\vert^2|\{\widehat{\mathbf{h}}_{i1}\}\right\}}{\left\Vert \argmin_{\{\mathbf{v}_{k1}\}} \quad \mathbb{E}\left\{\vert s_{k1} - \widehat{s}_{k1}\vert^2|\{\widehat{\mathbf{h}}_{i1}\}\right\}\right \Vert}\notag
\end{equation}
\begin{equation}
\qquad=\frac{\left(\sum_{i=1}^{K}p_i\left(\widehat{\mathbf{h}}_{i1}\widehat{\mathbf{h}}_{i1}^H + \widetilde{\mathbf{R}}_{i1}\right) + \sigma^2 \mathbf{I}_{N+1}\right)^{-1}\widehat{\mathbf{h}}_{k1}}{\left\Vert \left(\sum_{i=1}^{K}p_i\left(\widehat{\mathbf{h}}_{i1}\widehat{\mathbf{h}}_{i1}^H + \widetilde{\mathbf{R}}_{i1}\right) + \sigma^2 \mathbf{I}_{N+1}\right)^{-1}\widehat{\mathbf{h}}_{k1}\right\Vert}.
\end{equation}
For combining vectors $\mathbf{v}_{kl} \in \mathbb{C}^{(N+1)}, l = \{2,\cdots,L\}$ with $\Vert \mathbf{v}_{kl}\Vert^2 = 1$ we define following terms which would be useful for simplification of analysis: augmented channels, augmented channel estimates and augmented channel errors are required, which are given respectively as
\begin{equation}\label{def_c_ikl}
\mathbf{c}_{{i_k}l}\triangleq \begin{bmatrix}
\mathbf{h}_{il}\\
g_{{i_k}{(l-1)}}\end{bmatrix} ;~~\widehat{\mathbf{c}}_{{i_k}l}\triangleq \begin{bmatrix}
\widehat{\mathbf{h}}_{il}\\
\widehat{g}_{{i_k}{(l-1)}}\end{bmatrix} ;~~ \widetilde{\mathbf{c}}_{{i_k}l}\triangleq \begin{bmatrix}
\widetilde{\mathbf{h}}_{il}\\
\widetilde{g}_{{i_k}(l-1)}\end{bmatrix}
\end{equation}

\begin{equation}\label{Vk_NLMMSE}
\mathbf{v}_{kl}^{\mathrm{N-LMMSE}} = \frac{\argmin_{\{\mathbf{v}_{kl}\}} \quad \mathbb{E}\left\{\vert s_{kl} - \widehat{s}_{kl}\vert^2|\{\widehat{\mathbf{c}}_{j_kl}\}\right\}}{\left\Vert \argmin_{\{\mathbf{v}_{kl}\}} \quad \mathbb{E}\left\{\vert s_{kl} - \widehat{s}_{kl}\vert^2\{\widehat{\mathbf{c}}_{j_kl}\}\right\}\right \Vert}\notag
\end{equation}
\begin{equation}
\quad= \frac{\left(\sum_{i=1}^{K}p_i\mathbb{E}\left\{\mathbf{c}_{{i_k}l}\mathbf{c}_{{i_k}l}^H |\{\widehat{\mathbf{c}}_{j_kl}\}\right\} + \sigma^2 \mathbf{I}_{N+1}\right)^{-1}\mathbf{b}_{{i_k}l} }{\left\Vert \left(\sum_{i=1}^{K}p_i\mathbb{E}\left\{\mathbf{c}_{{i_k}l}\mathbf{c}_{{i_k}l}^H |\{\widehat{\mathbf{c}}_{j_kl}\}\right\} + \sigma^2 \mathbf{I}_{N+1}\right)^{-1}\mathbf{b}_{{i_k}l} \right\Vert}.
\end{equation}
where
\begin{equation}
\mathbf{b}_{{i_k}l}\triangleq\mathbb{E}\left\{\mathbf{c}_{{i_k}l}|\{\widehat{\mathbf{c}}_{j_kl}\}\right\}.\label{b_ikl}\notag
\end{equation}
The above expectation terms are given as
\begin{align}
\mathbb{E}\left\{\mathbf{c}_{{i_k}l}\mathbf{c}_{{i_k}l}^H |\{\widehat{\mathbf{c}}_{j_kl}\}\right\} &=\widehat{\mathbf{c}}_{{i_k}l}\widehat{\mathbf{c}}_{{i_k}l}^H + \begin{bmatrix}
\widetilde{\mathbf{R}}_{i{l}} &\mathbf{0}\label{Eq_CCh}\\
\mathbf{0}^T &\widetilde{\psi}_{{i_k}(l-1)}\end{bmatrix},\\
\mathbb{E}\left\{\mathbf{c}_{{i_k}l}|\{\widehat{\mathbf{c}}_{j_kl}\right\}&=\widehat{\mathbf{c}}_{{i_k}l}\label{Eq_C}.
\end{align}
Refer Appendix \ref{Expc_CCh} for computation of \eqref{Eq_CCh} and \eqref{Eq_C}.

\section{Simulation Results}

We consider the simulation setup illustrated in Fig.~\ref{fig0}. The APs are equally placed on a radio stripe of length $500\,\mathrm{m}$ which is wrapped around a square perimeter of the same length (e.g., factory, office, etc.,). For analysis to be simple we consider the following 3GPP Urban Microcell model \cite{bjrnson2019making} with $2$ GHz carrier frequency and

\begin{equation}
\beta_{kl} = -30.5 - 36.7\mathrm{log}_{10}\left(\frac{d_{kl}}{1\mathrm{m}}\right),
\end{equation}
where $d_{kl}$ is the distance between AP $l$ and UE $k$ which also includes the vertical height of $5\,\mathrm{m}$ difference between the APs and UEs. Each UE is assumed to transmit with $50$ mW power, the bandwidth is taken to be $20$ MHz, the noise power $\sigma^2$ is $-92$ dBm, the coherence block length $\tau_c$ is $ 200$ channel uses, and $\tau_p$ is $20$ orthogonal pilot sequences. The total number of APs is $L=24$ and each has $N=4$ antennas. The UEs are uniformly distributed within the square setup. The spatial correlation is modeled using the Gaussian  local scattering model \cite[Chapter 2]{mMIMObookEmil} unless otherwise stated.
\begin{figure}[t!]
	\centering
	\includegraphics[width=0.45\textwidth,trim={0.6cm 0.1cm 1.1cm 0.7cm},clip]{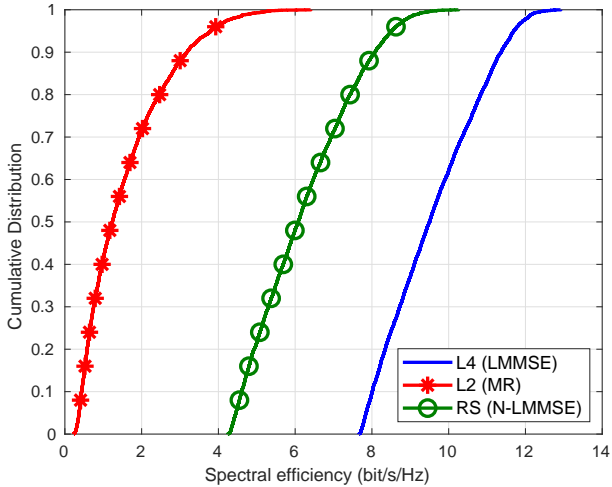}
	\vspace{-1mm}
	\caption{Comparison of performance of radio stripe, RS (N-LMMSE ) with L2 (MR) and cell-free L4 (LMMSE) processing ($L=24,\, K=10$).}
	\label{fig1} \vspace{-2mm}
\end{figure} 
In Fig.~\ref{fig1}, we compare the performance of the proposed sequential radio stripe (RS) using N-LMMSE processing with that of a centralized implementation of cell-free massive MIMO using LMMSE processing \cite{Nayebi2016a}, which is called level 4 (L4) processing in \cite{bjrnson2019making}. We also compare with conventional MR \cite{Hien,Nayebi2016a}, which can be treated as a level 2 (L2) processing in \cite{bjrnson2019making}. The figure shows the cumulative distributive function (CDF) of the SE of a randomly located UE, in the case of $K=10$. It can be observed that N-LMMSE significantly outperforms L2 (MR) processing. The superior performance of proposed method shows the importance of letting the APs cooperate in reducing interference. L4 processing has the highest performance since it is a centralized implementation where the CPU has access to the received signal of all the APs. This will require an immense front-haul signaling to implement. On other hand, in the proposed scheme the APs only cooperate pair-wise and strike a good trade-off between performance and front-haul signaling.
\begin{figure}[t!]
	\centering
	\includegraphics[width=0.45\textwidth,trim={0.6cm 0.1cm 1.1cm 0.7cm},clip]{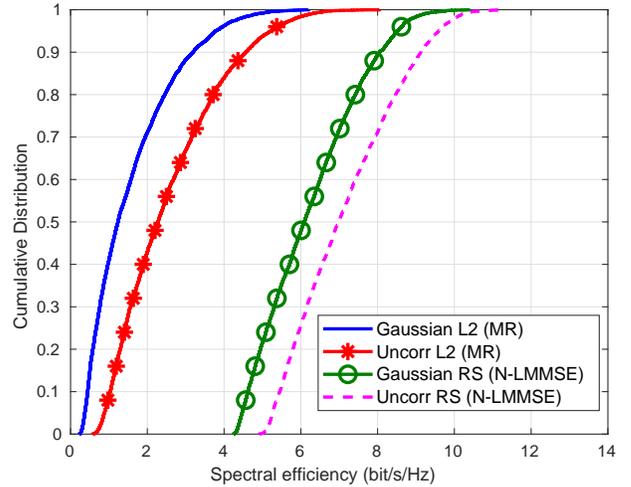}
		\vspace{-1mm}
		\caption{Comparison of performance of radio stripe with correlated and uncorrelated channel between APs and UEs  ($L=24,\, K=10$).}
		\label{fig2} \vspace{-5mm}
\end{figure} 
In Fig.~\ref{fig2}, we compare the performance of the proposed algorithm with spatially uncorrelated fading channels and the Gaussian local scattering model. It can be observed that uncorrelated channels have better performance. This is due to the similar spatial correlation matrices of the users in the correlated modeling since the nominal angles are similar \cite{mMIMObookEmil}. 
\begin{figure}[t!]
	\centering
	\includegraphics[width=0.45\textwidth,trim={0.6cm 0.1cm 1.1cm 0.7cm},clip]{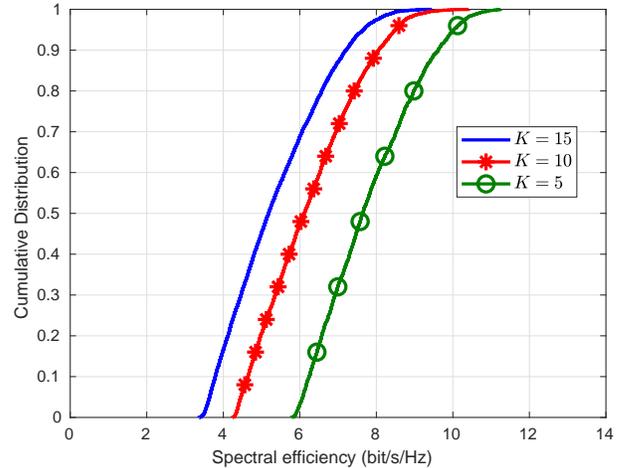}
		\caption{Comparison of performance of radio stripe when the number of UEs is varying with fixed $L=24$ and N-MMSE processing.}
		\label{fig3} \vspace{-4mm}
\end{figure} 
Fig.~\ref{fig3} shows the performance of the proposed method for a varying number of UEs. The SE reduces as $K$ increases, due to the additional interference. But the interference suppression of the sequential processing remains effective also with $K$ is large and comparable to $L$.

It can be noted with regard to front-haul signaling that
when using L4 implementation, $2NL\tau_c$ real-valued scalars must be sent to the CPU in every coherence block. With the proposed method, $3K^2+2K(\tau_c-\tau_p)$ real-valued scalars must be sent over every segment of the front-haul, including the one connected to the CPU. In the simulation setup considered, the proposed method reduces the front-haul signaling to the CPU by 90\% compared to L4 processing.
\vspace{-1mm}
\section{Conclusion}
\vspace{-1mm}
The paper introduces a sequential processing framework for interference suppression in cell-free mMIMO systems. It can be utilized when there is a sequential front-haul (e.g., implemented using a radio stripe architecture), where each AP is connected to two adjacent APs. We have derived the achievable SE using N-LMMSE combining. The proposed algorithm outperforms traditional MR and achieves a comparable performance to a fully centralized implementation of cell-free massive MIMO, while using 90\% less front-haul signaling in the simulation part. Hence, sequential processing finds a suitable trade-off between high SE and low cost and  front-haul requirements.
\vspace{-3mm}
\appendix
\section{Derivation of terms used}
\label{Expc_CCh}
\subsection{Derivation of $\mathbb{E}\left\{\mathbf{c}_{{i_k}l}\mathbf{c}_{{i_k}l}^H|\{\widehat{\mathbf{c}}_{j_kl}\}\right\}$ and $\mathbb{E}\left\{\mathbf{c}_{{i_k}l}|\{\widehat{\mathbf{c}}_{j_kl}\}\right\}$ terms in \eqref{Eq_CCh}, \eqref{Eq_C}}\label{Der_v_kl_LMMSE}
\vspace{-3mm}
\begin{equation}
\mathbf{c}_{{i_k}l}\mathbf{c}_{{i_k}l}^H = \begin{bmatrix}
\mathbf{h}_{il}\\
g_{{i_k}(l-1)}\end{bmatrix}\begin{bmatrix}
\mathbf{h}_{il}^H
g_{{i_k}(l-1)}^*\end{bmatrix}
\end{equation}
using \eqref{def_c_ikl} the above equation can be re-written as
\begin{flalign}
\begin{aligned}
\mathbf{c}_{{i_k}l}\mathbf{c}_{{i_k}l}^H =\widehat{\mathbf{c}}_{{i_k}l}\widehat{\mathbf{c}}_{{i_k}l}^H + \widehat{\mathbf{c}}_{{i_k}l}\widetilde{\mathbf{c}}_{{i_k}l}^H + \widetilde{\mathbf{c}}_{{i_k}l}\widehat{\mathbf{c}}_{{i_k}l}^H + \widetilde{\mathbf{c}}_{{i_k}l}\widetilde{\mathbf{c}}_{{i_k}l}^H
\end{aligned}
&&
\end{flalign}
and expectation over estimation errors conditioned on $ \{\widehat{\mathbf{c}}_{j_kl}:j=1,\cdots,K,~l=2,\cdots,L\}$ i.e., conditioned on $ \{\widehat{\mathbf{h}}_{jl}\},\{\widehat{g}_{{j_k}(l-1)}\}\, \forall j\in \{1,\cdots,K\}, \forall l\in \{2,\cdots,L\}$ is given by
\begin{flalign}
\begin{aligned}[b]
\mathbb{E}&\left\{\mathbf{c}_{{i_k}l}\mathbf{c}_{{i_k}l}^H|\{\widehat{\mathbf{c}}_{j_kl}\}\right\}\\ &= \mathbb{E}\left\{\widehat{\mathbf{c}}_{{i_k}l}\widehat{\mathbf{c}}_{{i_k}l}^H|\{\widehat{\mathbf{c}}_{j_kl}\}\right\} + \mathbb{E}\left\{\widehat{\mathbf{c}}_{{i_k}l}\widetilde{\mathbf{c}}_{{i_k}l}^H|\{\widehat{\mathbf{c}}_{j_kl}\}\right\} \\ &+ \mathbb{E}\left\{\widetilde{\mathbf{c}}_{{i_k}l}\widehat{\mathbf{c}}_{{i_k}l}^H|\{\widehat{\mathbf{c}}_{j_kl}\}\right\} + \mathbb{E}\left\{\widetilde{\mathbf{c}}_{{i_k}l}\widetilde{\mathbf{c}}_{{i_k}l}^H|\{\widehat{\mathbf{c}}_{j_kl}\}\right\},
\end{aligned}
&&
\end{flalign}
where,
\begin{flalign}
\begin{aligned}
\mathbb{E}&\left\{\widehat{\mathbf{c}}_{{i_k}l}\widehat{\mathbf{c}}_{{i_k}l}^H|\{\widehat{\mathbf{c}}_{j_kl}\}\right\} = \widehat{\mathbf{c}}_{{i_k}l}\widehat{\mathbf{c}}_{{i_k}l}^H\\&= \begin{bmatrix}
\widehat{\mathbf{h}}_{il}\widehat{\mathbf{h}}_{il}^H &\widehat{\mathbf{h}}_{il}\widehat{g}_{{i_k}(l-1)}^*\\
\widehat{g}_{{i_k}(l-1)}\widehat{\mathbf{h}}_{il}^H
&\vert \widehat{g}_{{i_k}(l-1)} \vert^2\end{bmatrix},
\end{aligned}
&&
\end{flalign}
\begin{flalign}
\begin{aligned}
\mathbb{E}&\left\{\widehat{\mathbf{c}}_{{i_k}l}\widetilde{\mathbf{c}}_{{i_k}l}^H|\{\widehat{\mathbf{c}}_{j_kl}\}\right\}=\widehat{\mathbf{c}}_{{i_k}l}\mathbb{E}\left\{\widetilde{\mathbf{c}}_{{i_k}l}^H|\{\widehat{\mathbf{c}}_{j_kl}\}\right\}\\&=\begin{bmatrix}
\widehat{\mathbf{h}}_{il}\\
\widehat{g}_{{i_k}{(l-1)}}\end{bmatrix}\begin{bmatrix}
\mathbf{0}^T &
0\end{bmatrix}=\mathbf{O}_{(N+1)},
\end{aligned}
&&
\end{flalign}
\begin{equation}
\mathbb{E}\left\{\widetilde{\mathbf{c}}_{{i_k}l}\widehat{\mathbf{c}}_{{i_k}l}^H|\{\widehat{\mathbf{c}}_{j_kl}\}\right\}=\mathbb{E}\left\{\widetilde{\mathbf{c}}_{{i_k}l}|\{\widehat{\mathbf{c}}_{j_kl}\}\right\}\widehat{\mathbf{c}}_{{i_k}l}^H=\mathbf{O}_{(N+1)},
\end{equation}
\begin{flalign}\label{var_ctilde}
\begin{aligned}
\mathbb{E}&\left\{\widetilde{\mathbf{c}}_{{i_k}l}\widetilde{\mathbf{c}}_{{i_k}l}^H|\{\widehat{\mathbf{c}}_{j_kl}\}\right\}\\&=\mathbb{E}\left\{ \begin{bmatrix}
\widetilde{\mathbf{h}}_{il}\widetilde{\mathbf{h}}_{il}^H &\widetilde{\mathbf{h}}_{il}\widetilde{g}_{{i_k}(l-1)}^*\\
\widetilde{g}_{{i_k}(l-1)}\widetilde{\mathbf{h}}_{il}^H
&\vert \widetilde{g}_{{i_k}(l-1)} \vert^2\end{bmatrix}\middle|\{\widehat{\mathbf{c}}_{j_kl}\}\right\}\\
&=\begin{bmatrix}
\widetilde{\mathbf{R}}_{i{l}} &\mathbf{0}\\
\mathbf{0}^T &\widetilde{\psi}_{{i_k}(l-1)}\end{bmatrix}.
\end{aligned}
&&
\end{flalign}
Its easy to observe for \eqref{Eq_C} it follows that
\begin{flalign}
\begin{aligned}
\mathbb{E}&\left\{\mathbf{c}_{{i_k}l}|\{\widehat{\mathbf{c}}_{j_kl}\}\right\}=\widehat{\mathbf{c}}_{{i_k}l}+\mathbb{E}\left\{\widetilde{\mathbf{c}}_{{i_k}l}|\{\widehat{\mathbf{c}}_{j_kl}\}\right\}=\begin{bmatrix}
\widehat{\mathbf{h}}_{il}\\
\widehat{g}_{{i_k}(l-1)}\end{bmatrix}.
\end{aligned}
&&
\end{flalign}
\bibliographystyle{IEEEtran}
\vspace{-2mm}
\bibliography{IEEEabrv.bib,reff.bib}

\end{document}